\documentclass{PoS}

\title{Construction of Triple-GEM Detector Using Commercially Manufactured Large GEM Foils}

\ShortTitle{Commercially Manufactured Large GEM Foils}

\author{\speaker{M. Posik}\\
        Temple University\\
        E-mail: \email{posik@temple.edu}}

\author{B. Surrow\\
        Temple University\\
        E-mail: \email{surrow@temle.edu}}

\abstract{Many experiments are currently using or proposing to use large area GEM foils in their detectors, which is creating a need for commercially available GEM foils. Currently CERN is the only main distributor of large GEM foils, however with the growing interest in GEM technology keeping up with the increasing demand for GEMs will be difficult. 

We present here an update on the assembly and testing of triple-GEM tracking detectors utilizing single-masked $40 \times 40$ cm$^2$ commercial GEM foils produced by Tech-Etch. The triple-GEM detectors will allow us to characterize the overall quality of these Tech-Etch foils through gain, efficiency, and energy resolution measurements. This will be done by constructing four single-mask triple-GEM detectors, using foils manufactured by Tech-Etch, which follow the design used by the STAR Forward GEM Tracker (FGT). The stack is formed by gluing the foils to the frames and then gluing the frames together. The stack also includes a Tech-Etch produced high voltage foil and a 2D $r-\phi$ readout foil. While one of the four triple-GEM detectors will be built identically to the STAR FGT, the other three will investigate ways in which to further decrease the material budget and increase the efficiency of the detector by incorporating perforated Kapton spacer rings rather than G10 spacing grids to reduce the dead area of the detector.  }

\FullConference{5th International Conference on Micro-Pattern Gas Detectors (MPGD2017)\\
		22-26 May, 2017\\
		Philadelphia, USA}

\begin{document}

\section{Introduction}
\begin{figure}[!h]
\centering
\includegraphics[width=0.5\columnwidth]{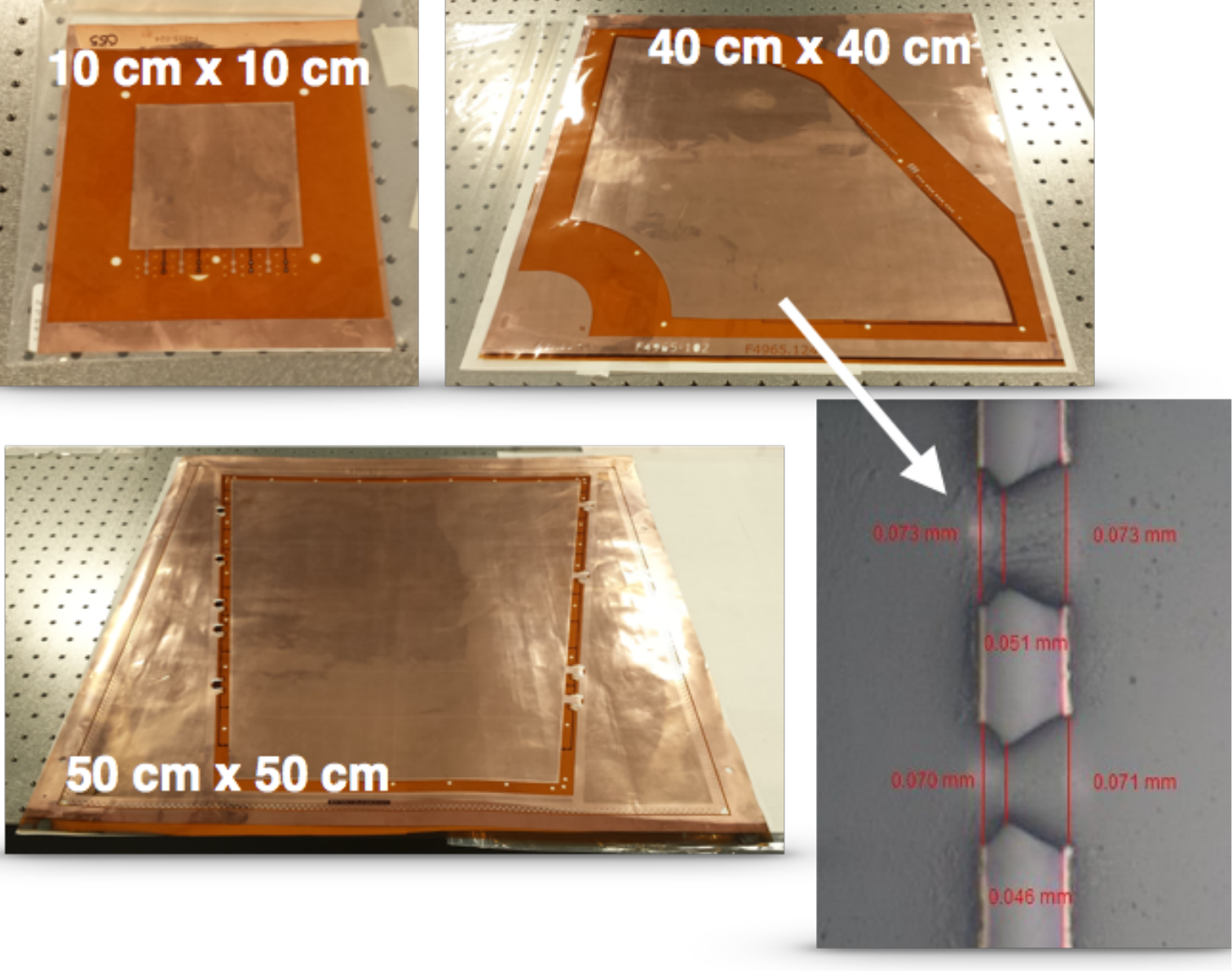}
\caption{Several different foils produced by Tech-Etch via the single-mask process. Upper left is a 10$\times$10 cm$^2$, upper right is a 40$\times$40 cm$^2$, and lower left is a 50$\times$50 cm$^2$ GEM foil. The lower right image is a cross sectional image showing the hole geometry for one of the 40$\times$40 cm$^2$ GEMs.}
\label{fig:foils}
\end{figure}

Many Experiments are currently using or are planning on using large area GEM based technologies for their detectors. However, with CERN being the main provider for large area GEM foils, there is a need for commercially available large GEM foils to help alleviate some of the demand. Tech-Etch, a US based company, acquired a license from CERN to produce GEM foils. Temple University then began working with Tech-Etch to help them establish a consistent GEM manufacturing procedure based on the single-mask technique, by providing fast quality checks of the GEM foils based on foil geometry and electrical performance. Before ending their GEM program, Tech-Etch had succeeded in producing single-mask GEM foils up to 50 cm $\times$ 50 cm in size. Figure ~\ref{fig:foils} shows a few of the different foils that Tech-Etch had produced, as well as a cross-sectional image of one of their GEMs.

\section{Design}

\begin{figure}[!h]
\centering
\includegraphics[width=0.75\columnwidth]{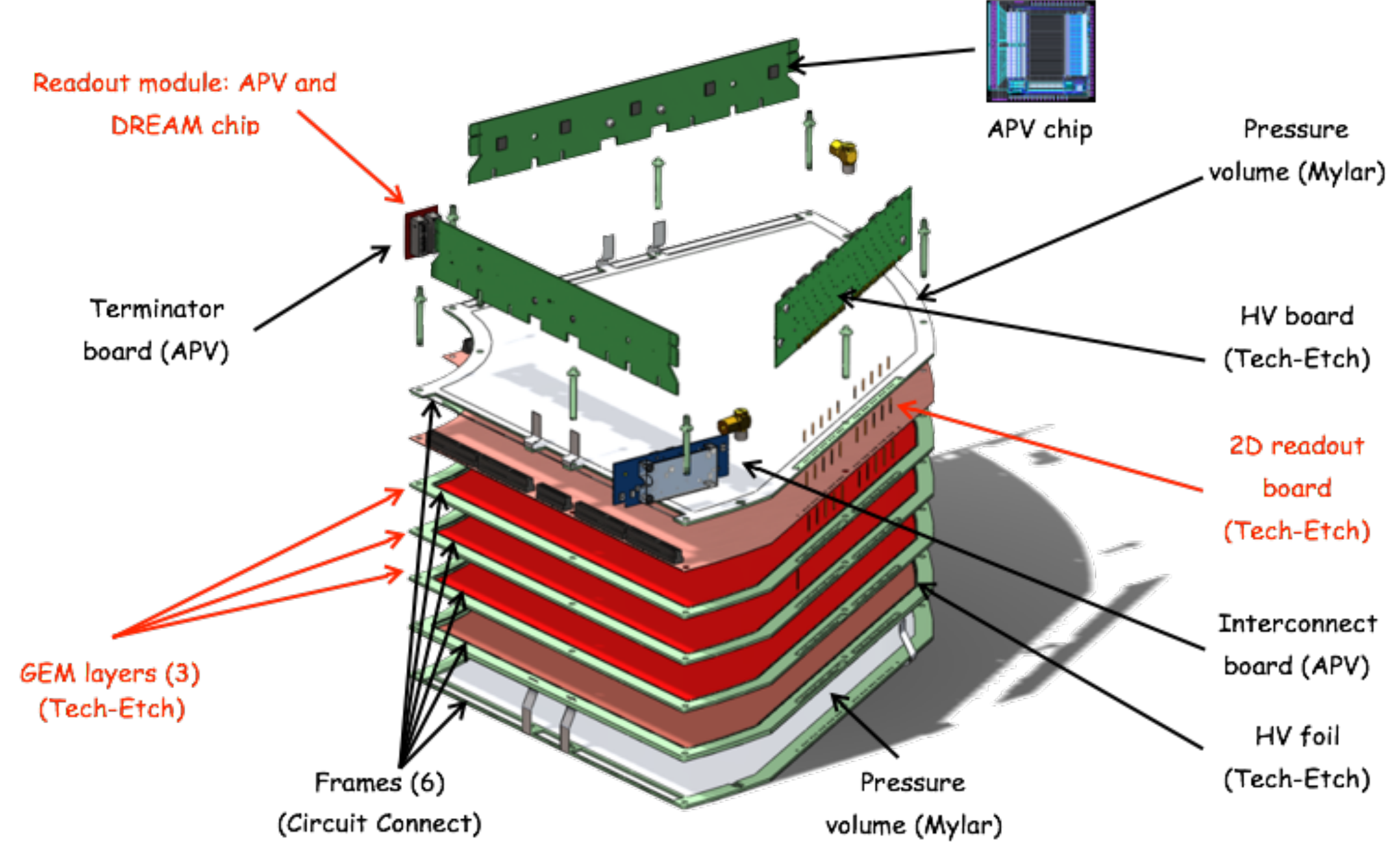}
\caption{Triple-GEM detector based on the STAR FGT~\cite{FGT} design, with the GEM foils to be replaced with Tech-Etch single-mask GEM foils.}
\label{fig:fgt}
\end{figure}

\begin{figure}[!h]
\centering
\includegraphics[width=0.75\columnwidth]{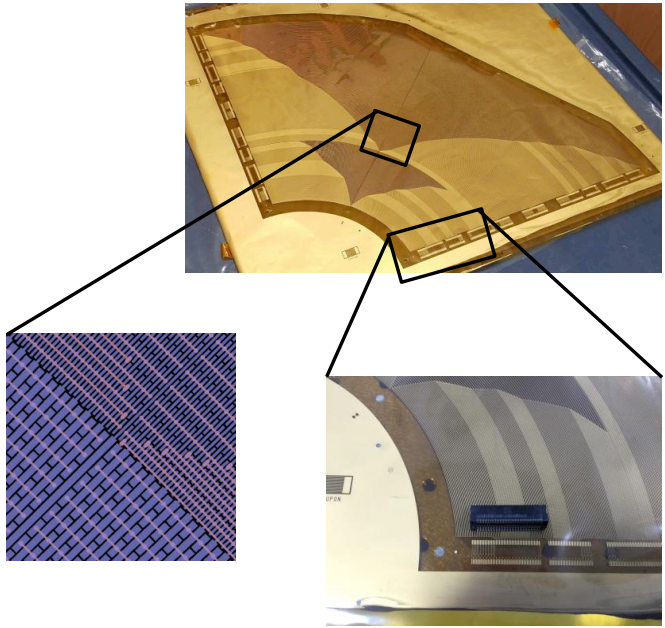}
\caption{2D r-$\phi$ readout foil produced by Tech-Etch. The pitch variation of r-$\phi$ readout strips can clearly be seen in the upper image. The lower left image is a design drawing of the variation near the center of the readout foil. The lower right image shows an unsoldered multi-pin connector near its solder points on the readout foil.}
\label{fig:readout}
\end{figure}

The Tech-Etch 40$\times$40 cm$^2$ single-mask GEM foils that have been optically analyzed for geometric properties and electrically tested via leakage current by Temple University showed comparable results to those obtained from CERN foils~\cite{NIMA-Posik}. The next step is to build several triple-GEM prototype detectors to measure and characterized the detector's efficiency, gain uniformity, and energy resolution. These measurements will provide the final verdict on the foil's quality. Temple University already has the tooling and DAQ setup that was used for the STAR Forward GEM Tracker (FGT)~\cite{FGT}, and so the prototypes will be based on this design, which is shown in Fig.~\ref{fig:fgt}. The triple-GEM detectors will use the Tech-Etch single-mask 40$\times$40 cm$^2$ GEM foils, which will be glued to frames, and the detector stack will be formed by gluing the frames, produced by Circuit Connect, together. The gas volume will then be sealed on the upper and lower side using metalized Mylar foil. The detector will follow a gap pattern of 3 (drift gap)/2/2/2 mm, respectively. Between each GEM layer we will place a G10 spacer-grid to maintain a consistent separation of the two GEM layers. Each GEM foil is divided into 9 high voltage (HV) segments, with each having an area of about 100 cm$^2$. The readout foil design is a 2D r-$\phi$ readout (Fig~\ref{fig:readout}), which was also produced by Tech-Etch.  The readout foil varies in pitch as a function of r and $\phi$ ranging from about 300-900 $\mu$m. The charge will be readout using 10 APV chips divided over two APV cards. The APV cards will connect to the readout foil through multi-pin connectors that were soldered to the readout foil by Proxy Manufacturing.

\begin{figure}[!h]
\centering
\includegraphics[width=0.75\columnwidth]{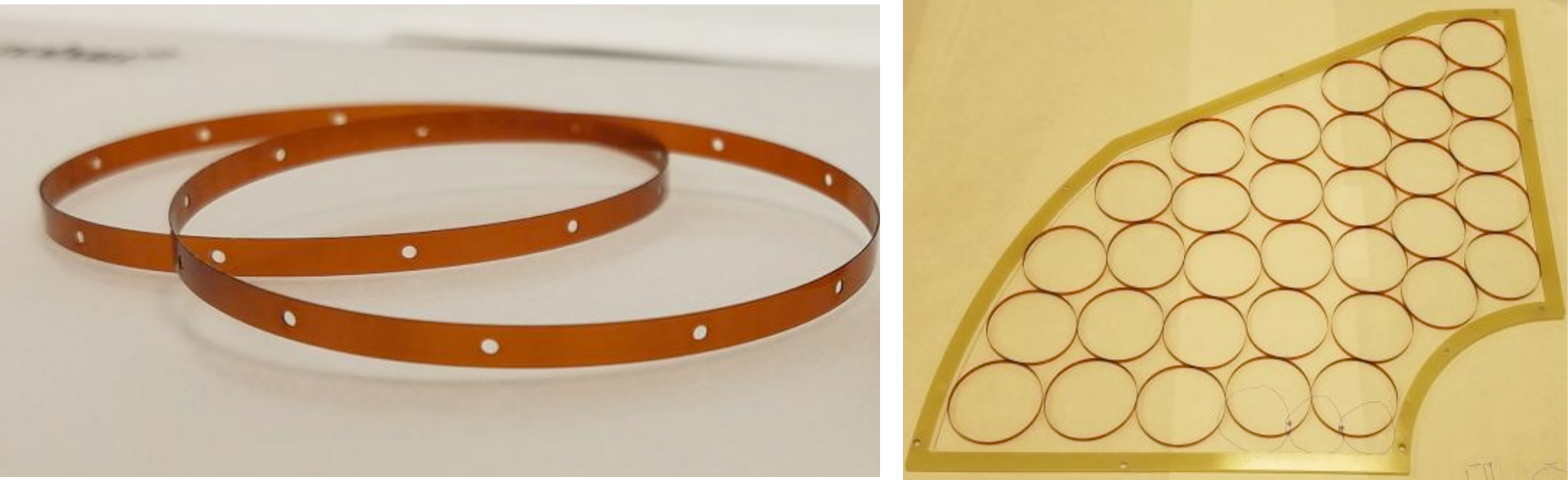}
\caption{Kapton spacer rings to be used in place of G10 spacer grids. The left image shows two Kapton spacer rings. The right image shows the packing arrangement used between each GEM layer.}
\label{fig:rings}
\end{figure}

In addition to testing the quality of the Tech-Etch foils using the traditional G10 spacer grids, we will also build a prototype which replaces the spacer grid layers with thin Kapton rings, shown in Fig.~\ref{fig:rings}, to try and reduce the detectors material budget. These Katpon rings, produced by Potomac Photonics, are perforated around their perimeter to allow gas flow through the detector, and have an inner diameter of about 50 mm, a wall thickness of 0.127 mm, and are cut into 2 and 3 mm lengths, to fill the respective gaps between GEM layers.

\section{Assembly} 
The prototype detectors will be built taking advantage of two main facilities at Temple University, a class 1000 cleanroom and a detector lab. All of the detector assembly where the inner volume of the detector is exposed to the environment will take place in the cleanroom. Once the detector has been assembled and sealed, it will be moved to our detector lab for further testing and characterization.

\begin{figure}[!h]
\centering
\includegraphics[width=0.75\columnwidth]{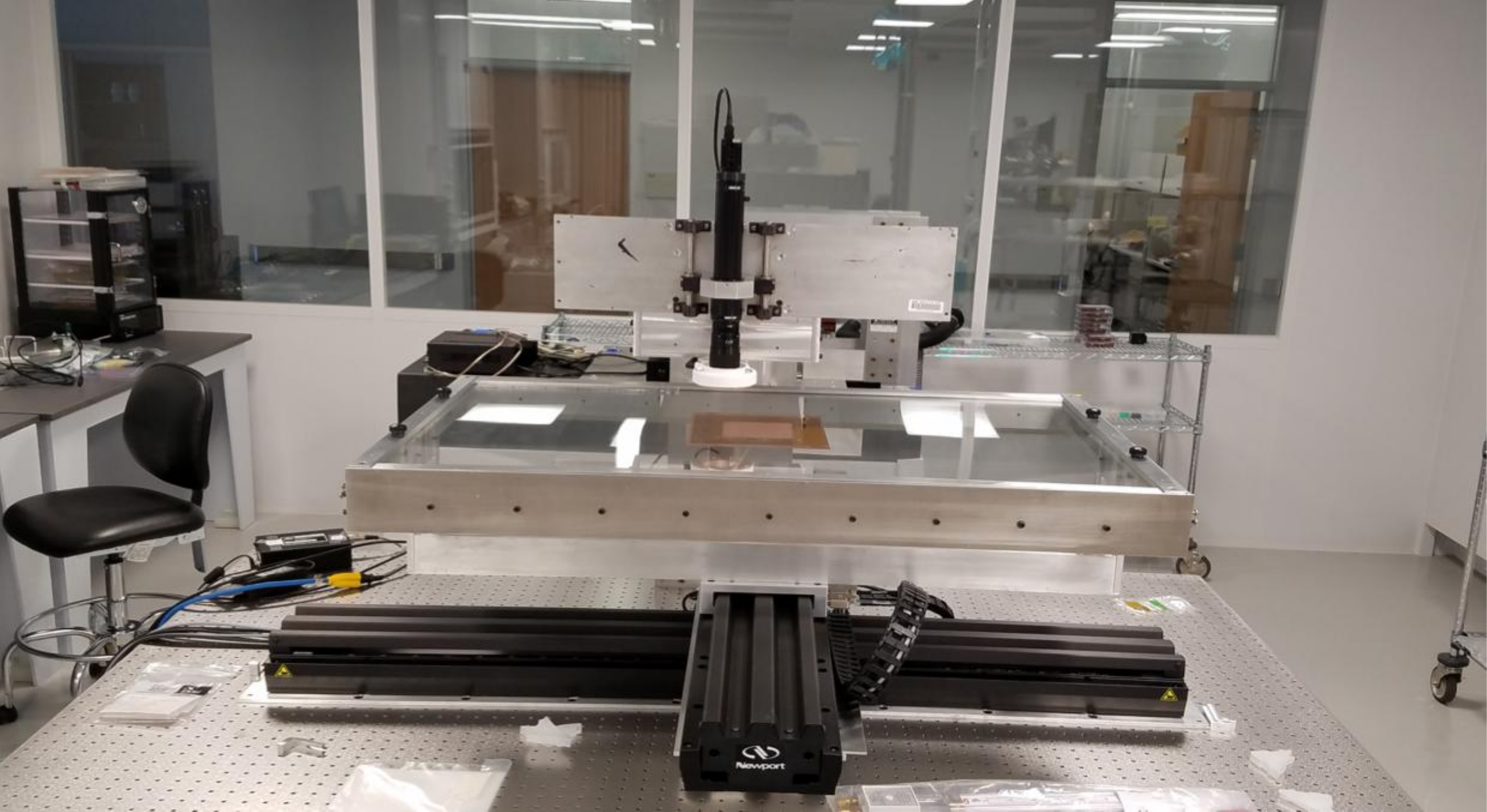}
\caption{CCD camera and 2D stage capable of scanning areas up to 100$\times$ 80 cm$^2$. This image shows a 10$\times$10 cm$^2$ GEM foil that is being scanned.}
\label{fig:ccd}
\end{figure}

The assembly process begins by lightly sanding and cleaning the frames which the GEM, readout, and HV foils will be glued to in an ultra-sonic bath with Di-water serving as the cleaning solution.   Once the frames are cleaned they are left to drip dry in the cleanroom for a couple days. In the cleanroom, the quality of the GEM foils that are to be used in the prototype detectors are accessed by first measuring their geometrical properties, such as inner and outer hole diameters, and their pitch. This is done with the use of a large area CCD scanner, capable of translating an area of up to 100$\times$ 80 cm$^2$, which images the GEM foils (Fig.~\ref{fig:ccd}). The images are then read into MATLAB where they are analyzed and their geometrical properties are determined. Next the bare GEM foil is placed in a N$_2$ environment where the leakage current of each of the nine sectors is measured with voltages of up to 500 V. During these measurements, the typical leakage current is around 1 nA~\cite{NIMA-Posik}. 

The GEM foil is then stretched using a pneumatic stretching tool, so that all wrinkles are removed from the GEM foil. The frame is then glued to the stretched GEM and is left to dry for 24 hours. The HV pins are then soldered onto their respective contact pads along the top of the GEM foil. After the HV pins have been soldered, the GEM layer is again placed into a N$_2$ environment and the leakage current of all sectors is measured. The HV, readout, and remaining two GEM foils follow a similar stretching, gluing, and testing procedure. Additionally, the top and bottom frames also have a metalized Mylar foil stretched and glued to it, which acts to seal the detector gas volume. Once the foils are glued to their respective frames, the detector stack is formed by gluing the frames from the various layers together such that the stack has gaps of 3/2/2/2 mm, where the 3 mm gap is between the GEM foil and the HV foil. During the stacking, spacers will be inserted in between each of the GEM foil layers. We will be testing two prototypes, one where thin G10 spacer grids are used, and another where the Kapton spacer rings are used.

\begin{figure}[!h]
\centering
\includegraphics[width=0.5\columnwidth]{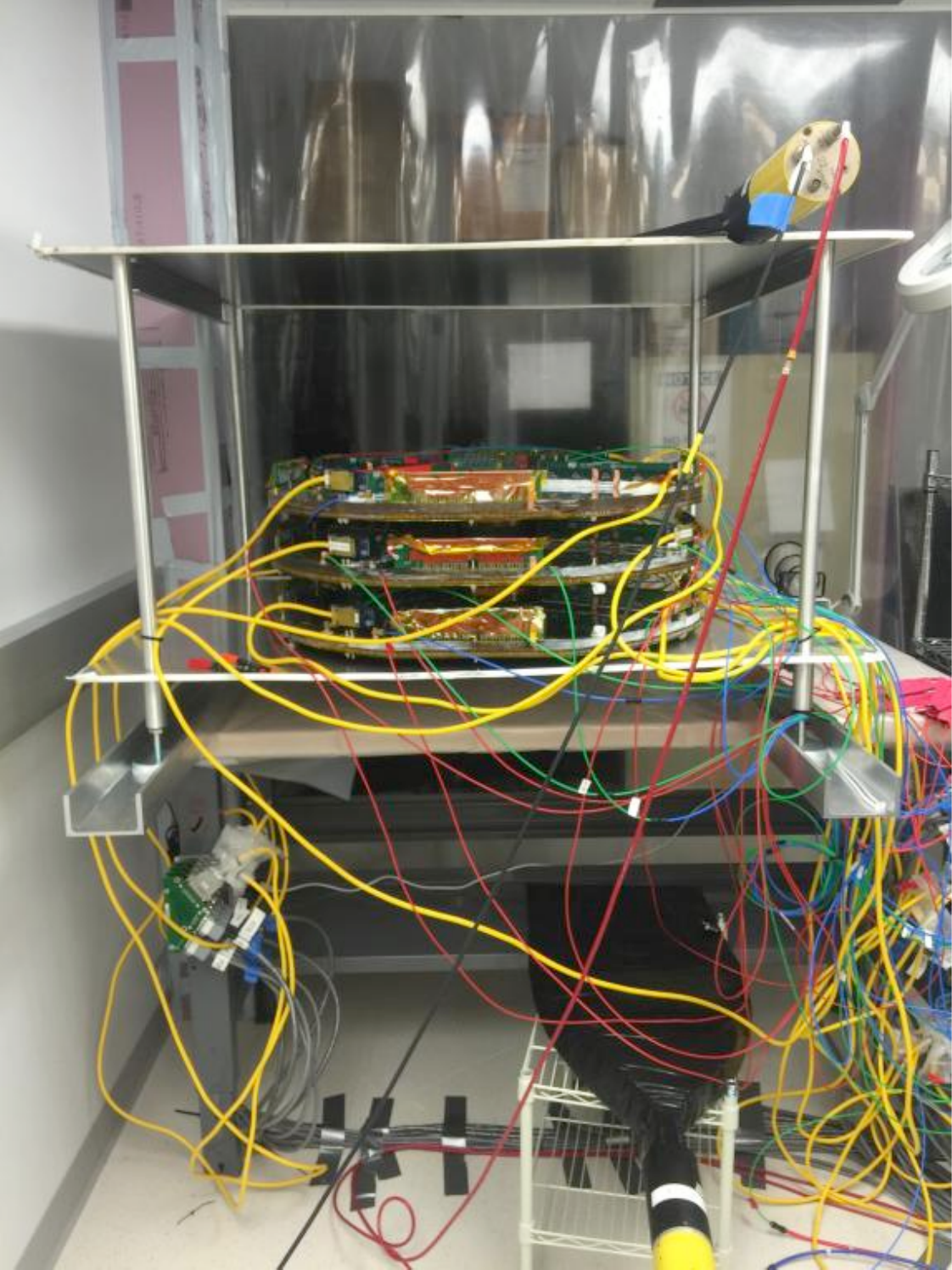}
\caption{Cosmic ray bench located in our detector lab.}
\label{fig:cosmics}
\end{figure}

Once the detector is sealed it is brought into the detector lab, where the HV distribution board, APV cards, and gas lines are connected to the detector. The detector is then made gas tight by flowing ArCO$_2$ through it and leak checking for CO$_2$. Once no more CO$_2$ leaks are detected the leakage current of the detector is checked again. Once the detector passes the leakage current test it is incorporated into our cosmic ray test stand, shown in Fig.~\ref{fig:cosmics}, to characterize the detector's efficiency. 

The final detector characterization would then be to measure its energy resolution and gain performance through the use of an $^{55}$Fe source.   

\section{Summary}
The successful electrical and geometrical tests of commercial GEM foils produced by Tech-Etch~\cite{NIMA-Posik}, has led us to build prototype triple-GEM detectors with critical components begin developed by commercial manufacturers (Tech-Etch), including the GEM, HV, and readout foils. These prototype detectors will allow us to better access the quality of the commercial GEM foils by quantifying the detectors efficiency, energy resolution, and gain. In addition to further testing the quality of the Tech-Etch GEM foils, we will also investigate replacing the traditional G10 spacer grids, used between GEM layers of a triple-GEM detector, by Kapton spacer rings to help reduce the overall detector material.  All materials and tooling needed to build several prototype triple-GEM detectors have now been acquired. As of this proceeding we have built two triple-GEM detectors, one using the G10 space grids and another using the Kapton spacer rings. These detectors are now being prepared for testing and characterization via cosmics.

\end{document}